# Nernst effect in single crystals of the pnictide superconductor CaFe$_{1.92}$Co$_{0.08}$As$_2$ and parent compound CaFe$_2$As$_2$

Marcin Matusiak[1,*], Zbigniew Bukowski[2], and Janusz Karpinski[2]

1. Institute of Low Temperature and Structure Research, Polish Academy of Sciences,
P.O. Box 1410, 50-950 Wrocław, Poland

2. Laboratory for Solid State Physics, ETH Zurich, 8093 Zurich, Switzerland

We report a combined study of the Nernst coefficient ($\nu$), Hall effect and thermoelectric power of CaFe$_2$As$_2$ and CaFe$_{1.92}$Co$_{0.08}$As$_2$ single crystals. The absolute value of $\nu$ in both samples is enhanced, probably due to ambipolar flow of electron- and hole-like quasiparticles. The onset of spin-density-wave order in CaFe$_2$As$_2$ causes further rapid rise of the Nernst coefficient. On the contrary, in the CaFe$_{1.92}$Co$_{0.08}$As$_2$ crystal we do not see any feature of $\nu$(T), which could be clearly associated with SDW fluctuations. In this Co-doped sample there is also no noticeable increase in of $\nu$ in the vicinity of superconducting transition, despite the expectation of such due to vortex movement.

[*] Corresponding author. Tel.: +48-71-3435021; Fax: +48-71-3441029; e-mail: M.Matusiak@int.pan.wroc.pl

The Nernst effect has become a subject of great interest, since it turned out to be a sensitive probe of exotic properties of the electronic matter [1,2,3]. It has been found to be of great use also in studying the normal [4,5] and mixed state [1] of superconductors. The recently discovered superconducting iron-pnictides seem to be no exception here. For instance, data for the Nernst effect in $LaO_{1-x}F_xFeAs$ [6] might suggest the presence of spin-density-wave (SDW) fluctuations near the superconducting transition. In the present paper we investigate properties of two samples belonging to the "122" family of the Fe-pnictides. Namely, we study transport phenomena, including the Nernst effect, in two single crystals: the $CaFe_2As_2$ parent compound and the cobalt-doped $CaFe_{1.92}Co_{0.08}As_2$ superconductor.

Single crystals of $CaFe_{2-x}Co_xAs_2$ were grown from Sn flux, using atomic ratios Ca:Fe:As:Sn=1.1:2:2.1:40 for pure $CaFe_2As_2$ and Ca:Fe:Co:As:Sn=1.1:1.85:0.2:2:40 for the Co-substituted compound. The components were loaded into alumina crucibles and placed in quartz ampoules which were evacuated and backfilled with Ar gas under pressure of 0.3 bar and sealed. The ampoules were heated to 600° C over 4 h, held there for 1 h, then heated in 5 h to 1050° C and kept at that temperature for 5 h so that all the components dissolved in the Sn flux. The ampoules were then cooled slowly at 2 K/h down to 600° C followed by furnace cooling to room temperature. The flux was removed by etching in diluted hydrochloric acid. Plate-like single crystals of typical dimensions 5 x 4 x 0.2 mm$^3$ with the c-axis perpendicular to the plate were obtained. However, those chosen for further investigation were smaller – of size ~ 2 x 1.5 x 0.1 mm$^3$. Phase purity was checked by powder X-ray diffraction (XRD). All the observed diffraction lines on the XRD pattern could be indexed according to the $I4/mmm$ space group. The lattice parameters obtained for $CaFe_2As_2$: $a = 3.883(1)$ Å and $c = 11.725(5)$ Å agree well with the values reported by Ronning at al. [7]. The lattice parameters $a = 3.886(1)$ Å and $c = 11.695(5)$ Å obtained for $CaFe_{1.92}Co_{0.08}As_2$ are close to those reported for $CaFe_{1.94}Co_{0.06}As_2$ [8]. A chemical composition of the Co-doped single crystal was determined by the energy dispersive x-ray (EDX) analysis.

The resistivity ($\rho$) was measured by the four-probe technique, with 25 $\mu$m gold wires attached to the crystal using two component silver epoxy. For the Hall coefficient ($R_H$) measurement, a sample was mounted on a rotatable probe and continuously turned by 180 degree (face down and up) in a magnetic field ($B$) of 13 T, to effectively reverse the field anti-symmetrical signal. During the thermoelectric power ($S$) and Nernst coefficient ($\nu$) measurements crystals were clamped between two phosphor bronze blocks, to what two Cernox thermometers and resistive heaters were attached. These heaters were cyclically switched on and off and the temperature runs were repeated at various values of the magnetic field (from -13 to +13 T) in order to extract the field voltage components, that were odd and even in $B$.

The temperature dependences of the resistivity for the $CaFe_2As_2$ and $CaFe_{1.92}Co_{0.08}As_2$ single crystals are shown in Fig. 1.

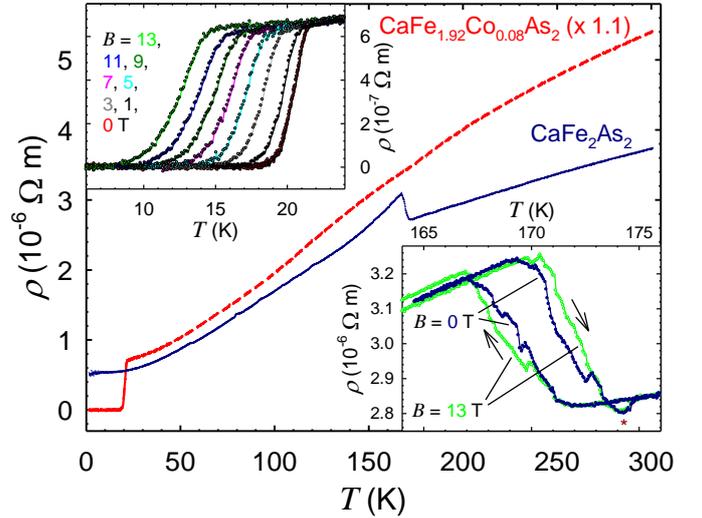

FIG. 1. In-plane resistivity versus temperature for $CaFe_2As_2$ (dark blue line) and $CaFe_{1.92}Co_{0.08}As_2$ (dashed red line) single crystals. The resistivity of the latter is multiplied by a factor of 1.1 for clarity. The upper inset shows $\rho(T)$ data for $CaFe_{1.92}Co_{0.08}As_2$ in the vicinity of superconducting transition. The applied magnetic field of 0, 1, 3, 5, 7, 9, 11, 13 T was parallel to the crystallographic $c$ axis. The lower inset shows $\rho(T)$ data for $CaFe_2As_2$ near $T_{tr}$. They were taken on warming and cooling (arrows indicate directions) and in the magnetic field of 0 T (solid dark blue points), and 13 T (open light green points), where $B \perp c$.

Our results remain in a good agreement with data presented by N. Kumar and coworkers [8], but it has to be mentioned that the reported value of the room temperature resistivity for $CaFe_2As_2$ single crystals varies among different studies. For example, $\rho$ (300K) was shown to be equal to: 2.2 $\mu\Omega$ m [7], 2.8 $\mu\Omega$ m [8], 3.1 $\mu\Omega$ m [9], 3.7 $\mu\Omega$ m (this report) and 5.6 $\mu\Omega$ m [10]. It is not yet clear what causes these differences. The pure $CaFe_2As_2$ single crystal undergoes the first order structural/magnetic transition at $T_{tr}$ = 170 K [11]. There can be seen a clear hysteresis in the resistivity ($\Delta T$ ~ 3 K) that is shown in the lower inset in Fig. 1 (data were collected at the heating/cooling rate of 0.1 K/min). The magnetic field of 13 T, which in our setup was perpendicular to the crystallographic $c$ axis, has no noticeable impact on the shape and position of the anomaly, whereas steps in region of the transition should probably be attributed to movements of the domain walls that has been recently detected in this material [12]. We are not certain at the moment, what is an origin of the dip at $T$ = 174.2 K (marked with asterisk), which appears only when the temperature is rising. There is no sign of a structural or spin-density-wave (SDW) transition in the $\rho$ (T) dependency for $CaFe_{1.92}Co_{0.08}As_2$. The sharp superconducting transition ($\Delta T_{10-90\%}$ ~ 2 K) occurs in the absence of magnetic field ($B$) at $T_c$ = 20.4 K (the middle point). An application of $B$ parallel to the crystallographic $c$ axis shifts $T_c$ to lower temperatures (see the upper inset in Fig. 1) and it reaches $T_c$ = 12.6 K at $B$ = 13 T. The resistive superconducting transitions are only slightly broadened by the magnetic field, and the $\rho$ (T,B) curves do not show the "fanning out" property, which is typical of the cuprates.

The temperature dependences of the Hall coefficients for the $CaFe_2As_2$ and $CaFe_{1.92}Co_{0.08}As_2$ crystals are presented in the upper panel (a) of Fig. 2. Both $R_H(T)$ values are negative, except for a region of 40 K below $T_{tr}$, where $R_H$ in the unsubstituted crystal becomes positive. This may be related to the reconstruction of the Fermi surface [13] resulting from the structural/magnetic transition. On the other hand, $CaFe_2As_2$ is a conductor with the multi-component Fermi surface with charge carriers of the opposite signs [14]. Therefore, a change of the $R_H$ sign can result from a drastic modification of the hole/electron mean-free paths, as suggested at the transition [15]. The importance of this mechanism seems to be supported by data on the Hall coefficient in $CaFe_2As_2$ reported by F. Ronning and coworkers in Ref. [7]. They noticed a very similar hump in $R_H(T)$ below $T_{tr}$, but there no change of sign was observed.

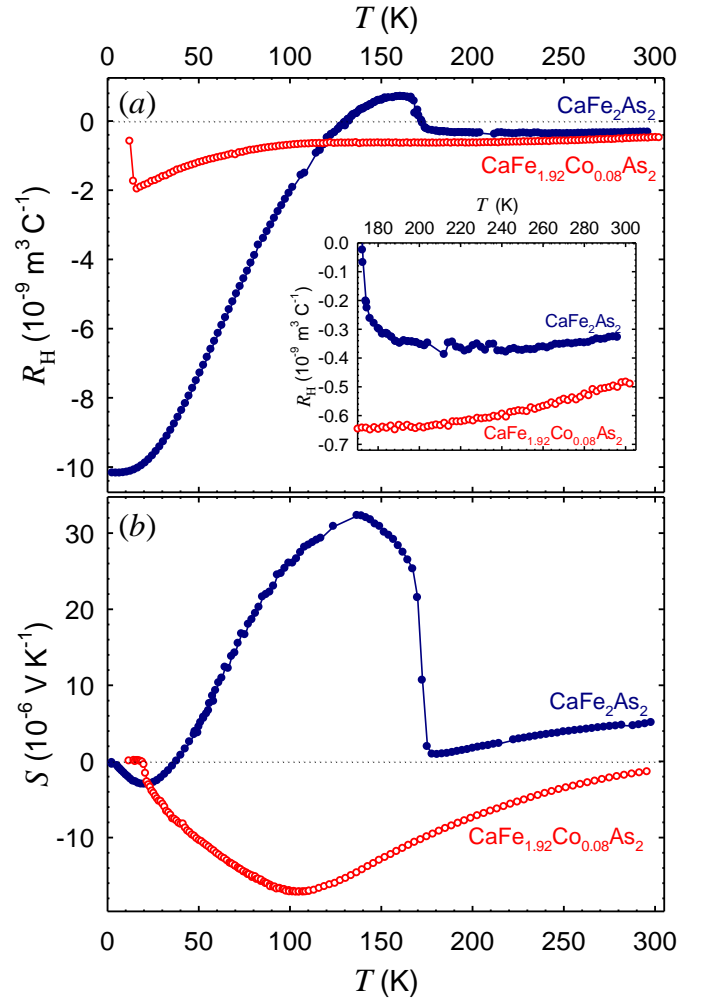

FIG. 2. (a) Temperature dependences of the Hall coefficient for $CaFe_2As_2$ (solid dark blue points) and $CaFe_{1.92}Co_{0.08}As_2$ (open red points) single crystals. The inset shows enlarged region above $T_{tr}$. (b) Temperature dependences of the thermoelectric power for $CaFe_2As_2$ (solid dark blue points) and $CaFe_{1.92}Co_{0.08}As_2$ (open red points) single crystals.

The total Hall coefficient of a multi bands conductor is a sum of the Hall coefficients of individual bands (denoted by "+" and "-" indexes), weighted by square of their electrical conductivities ($\sigma$):

$$R_H = \frac{R_{H+}(\sigma_+)^2 + R_{H-}(\sigma_-)^2}{(\sigma_+ + \sigma_-)^2}. \quad (1)$$

Therefore, in two single crystals with different number of scattering impurities, which can affect $\sigma_+$ and $\sigma_-$ in different ways, $R_H$ may have different sign in a certain range of $T$. An analogous variation of the Hall coefficient can be observed in the Co-doped crystal for temperatures above $T_{tr}$. In this region $R_H(T)$ is shifted vertically to more negative values as compared to the pure sample (see the inset in Fig. 2(a)). A ratio of $R_H$ for $CaFe_{1.92}Co_{0.08}As_2$ to $R_H$ for $CaFe_2As_2$ is at room temperature $\approx 1.5$, thus a rise of the number of the electron-like charge carriers would have to be significant to induce such a big effect. On the contrary, the resistivity data do not show reduction but growth of the high temperature resistivity in the Co-doped crystal, what suggests that the observed variation of the Hall coefficient may result rather from a decrease of the holes mean-free path.

This scenario seems to be proved by the high temperature data on the thermoelectric power shown in Fig. 2(b). This quantity can be also expressed as a sum of contributions from hole-like and electron-like carriers, but with different conductivity weighting factors:

$$S = \frac{S_+\sigma_+ + S_-\sigma_-}{\sigma_+ + \sigma_-}. \quad (2)$$

In the $CaFe_{1.92}Co_{0.08}As_2$ crystal $S(T)$ above $T_{tr}$ is vertically shifted from positive to negative values. Because this shift is almost parallel, we conclude that it arises from $T$-independent change in the electrical conductivities, and it is likely to be caused by the additional scattering of the hole-like charge carriers caused by cobalt ions. Below $T_{tr}$ the thermopower in pure $CaFe_2As_2$ develops a high and broad maximum that is one of characteristics of spin-density-wave order [16] and it was already observed in this material [10].

As shown in Fig. 3 the temperature dependence of the Nernst coefficient in pure $CaFe_2As_2$ also shows a large anomaly below $T_{tr}$. This rapid growth of the absolute value of $\nu$ has likely the same origin as the maximum in $S(T)$, i.e. it can be caused by the transition of the electronic system to the SDW state. It has been recently shown by A. Hackl and S. Sachdev that a reconstruction of the Fermi surface due to spin-density-wave order has fundamental implications for the Nernst signal and the thermopower [17]. The authors relate this dramatic change of the thermoelectric response to the singularity in the quasiparticle density of states. A different approach utilized by V. Oganesyan and I. Ussishkin also indicates a possible significant enhancement of the Nernst signal in the case of formation of Fermi pockets associated with spin-density-waves [18].

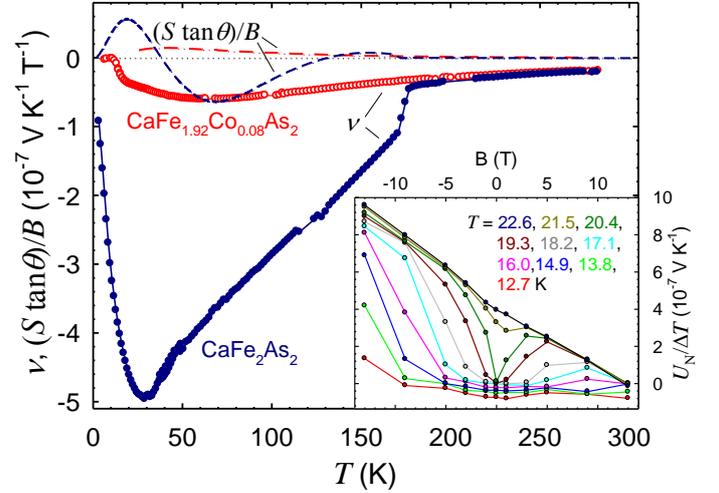

FIG. 3. Temperature dependences of the Nernst coefficient and $S\tan\theta$ term of the $CaFe_2As_2$ (solid dark blue points, and dashed dark blue line respectively) and $CaFe_{1.92}Co_{0.08}As_2$ (open red points, and dash-dot red line) single crystals. The inset shows the transverse voltage divided by value of the longitudinal temperature difference as a function of the magnetic field.

In general, the large contribution to the thermoelectric power, which is seen below SDW transition, can be interpreted as a consequence of variation of the chemical potential [19] that is immediately connected with the Nernst coefficient [20]. Namely, $\nu$ is related to the Fermi temperature ($T_F$) and Hall mobility ($\mu_H$) by equation:

$$\nu = \frac{\pi^2 k_B}{3e}\frac{T\mu_H}{T_F}, \quad (3)$$

where $k_B$ is the Boltzmann constant, and $e$ is the elementary charge. The low temperature limits of $\mu_H \approx 0.02$ $T^{-1}$ and $\nu/T \approx 2.5\times 10^{-8}$ V K$^{-2}$ T$^{-1}$ yields the very small Fermi temperature in $CaFe_2As_2$: $T_F \approx 230$ K. This value is close to those deducted from the Hall effect analysis for the Sr-122 and Ba-122 parent compounds [9], but it should be treated only as an approximation, since $CaFe_2As_2$ is a multi-band system and the Fermi energy and mobility can vary among the bands.

The normal state Nernst coefficient is composed of two terms [21]:

$$\nu = \left(\frac{\alpha_{xy}}{\sigma} - S\tan\theta\right)\frac{1}{B}, \qquad (4)$$

where $\alpha_{xy}$ is the off-diagonal element of the Peltier tensor, and $\theta$ is the Hall angle. For a momentum and energy independent relaxation time, two terms in Eq. 4 cancel exactly [22]. This "Sondheimer cancellation" is no longer valid, when there are separate electron and hole conductivity bands [23], and this effectively leads to substantial enhancement of the measured Nernst signal. As shown in Fig. 3 the ($S\tan\theta$) component for $CaFe_2As_2$ and $CaFe_{1.92}Co_{0.08}As_2$ is much smaller than $\nu$ (thus $\frac{\alpha_{xy}}{\sigma} \gg S\tan\theta$) in the whole temperature range of study. It indicates that the Sondheimer cancellation is violated and suggests that simultaneous flow of electron-like and hole-like carriers occurs both inside and outside the SDW phase.

The $\nu(T)$ dependence in the normal state of $CaFe_{1.92}Co_{0.08}As_2$ does not show any particular feature, which could be attributed to the influence of the SDW fluctuations. The Nernst coefficient evolves smoothly and a shape of the $\nu(T)$ curve resembles closely $S(T)$. The main difference is the temperature, at which a broad minimum occurs. It is $T \approx 65$ K for the Nernst coefficient, and $T \approx 105$ K for the thermopower. In the vicinity of the superconducting transition in $CaFe_{1.92}Co_{0.08}As_2$, we do not see any Nernst effect contribution from vortex movement ($\nu_s$). A similar phenomenon was already observed in classical superconductors [24], where the "vortex liquid" state is extremely narrow, and also in $CeCoIn_5$ [3]. However, in this latter case it was possible to extract the flux flow signal with the aid of the thermopower data, and $\nu_s$ was estimated to be ~ 0.1 $\mu$V K$^{-1}$ T$^{-1}$. In our experiment the large value of the Nernst coefficient in the normal state can also make it difficult to extract a vortex contribution, but we estimate that value of $\nu_s$ would have to be smaller than ~ 0.01 $\mu$V K$^{-1}$ T$^{-1}$ and/or be present in a very narrow temperature range (~ 1 K) to remain undetected.

Furthermore, the field dependences of the Nernst signal for several temperatures close to $T_c$, which are shown in the inset in Fig. 3, do not show up any detectable positive contribution from vortex motion. The non-linear contribution seen for 16 K $\leq T \leq$ 20.5 K appears to be simple consequence of the superconducting state being destroyed by the increasing magnetic field. It could be mistaken as a component from flux-flow, if the experiment were carried out only for positive magnetic field. However, data taken for $B < 0$ reveal a gradual reappearance of the normal state behavior.

We think that the lack of a detectable contribution to the Nernst signal from flux flow can be a result of strong vortex pinning in the mixed state of $CaFe_{1.92}Co_{0.08}As_2$, which prevents vortices from flowing under the influence of the thermal gradient. This conclusion agrees with results of high-field magnetotransport and magnetization studies on a single crystal of the $Ba(Fe_{1-x}Co_x)_2As_2$ that indicate weak thermal fluctuations and/or strong vortex pinning in this material [25]. Results of magnetization measurements performed for the $CaFe_{1.92}Co_{0.08}As_2$ single crystal show that the irreversibility field is very close to the upper critical field and these data will be presented in a separate report.

In summary, we measured the transport coefficients of the parent compound $CaFe_2As_2$ as well as the superconducting $CaFe_{1.92}Co_{0.08}As_2$ single crystals. We found the Nernst coefficient in both samples enhanced due to ambipolar flow of quasiparticles. Additionally, the absolute value of $\nu$ grows rapidly and almost linearly with $T$ in the SDW state of $CaFe_2As_2$. The low-temperature values of $\nu$, $R_H$ and $\rho$ in this parent compound allow us to estimate the Fermi energy to be of the order of 20 meV. We did not see any clear evidence of SDW fluctuations in any measured quantity for $CaFe_{1.92}Co_{0.08}As_2$. In this superconducting material we also did not see any Nernst signal from vortex motion in the vicinity of $T_c$. It could mean that a vortex pinning in the Co-doped sample is very high and suppresses formation of the "vortex liquid" state.

*Acknowledgements*

The authors are grateful to J.R. Cooper for useful comments on this work and to M. Małecka for performing the EDX analysis.